\documentclass[12pt,a4,epsf]{article}

\usepackage{amsmath}
\usepackage{amssymb}
\makeatletter
\@addtoreset{equation}{section}
\makeatother

\newcommand{\VEV}[1]{\left\langle #1 \right\rangle}

\newcommand{\nn}{\nonumber}
\newcommand{\diag}{{\rm diag}}
\newcommand{\ie}{{\it i.e.}}
\newcommand{\eg}{{\it e.g.}}

\newcommand{\Mu}{M_{\mrm {GUT}}}
\newcommand{\MW}{M_{\mrm W}}

\newcommand{\MSB}{M_{\mrm {SB}}}
\newcommand{\order}[1]{${\cal O}(#1)$}

\newcommand{\GeV}{\mbox{GeV}}

\newcommand{\Gsm}{$SU(3)$\sub C$\times SU(2)$\sub L$\times U(1)$\sub Y}
\newcommand{\Ga}{$SU(3)$\sub C$\times SU(2)$\sub L$\times SU(2)$\sub R
                 $\times U(1)_{B-L}$}
\newcommand{\E}[1]{$E_#1$}

\newcommand{\abs}[1]{\left| #1 \right|}
\newcommand{\s}[1]{\widetilde{#1}}

\newcommand{\cc}[1]{\overline{#1}}
\newcommand{\sub}[1]{$_{\mrm{#1}}$}

\newcommand{\bequ}{\begin{equation}}
\newcommand{\eequ}{\end{equation}}
\newcommand{\beqn}{\begin{eqnarray}}
\newcommand{\eeqn}{\end{eqnarray}}
\newcommand{\bctr}{\begin{center}}
\newcommand{\ectr}{\end{center}}
\newcommand{\bit}{\begin{itemize}}
\newcommand{\eit}{\end{itemize}}
\newcommand{\Ls}{\left(}
\newcommand{\Rs}{\right)}

\newcommand{\hsp}[1]{\hspace {#1cm}}
\newcommand{\vsp}[1]{\vspace {#1cm}}

\newcommand{\mrm}{\rm}

\begin{document}
\begin{titlepage}

\begin{flushright}
hep-ph/0305116\\
KUNS-1843\\
\today
\end{flushright}

\vspace{4ex}

\begin{center}
{\large \bf
Sliding Singlet Mechanism Revisited
}

\vspace{6ex}

\renewcommand{\thefootnote}{\alph{footnote}}
Nobuhiro Maekawa\footnote
{e-mail: maekawa@gauge.scphys.kyoto-u.ac.jp
}
and 
Toshifumi Yamashita\footnote{
e-mail: yamasita@gauge.scphys.kyoto-u.ac.jp
}

\vspace{4ex}
{\it Department of Physics, Kyoto University, Kyoto 606-8502, Japan}\\

\end{center}

\renewcommand{\thefootnote}{\arabic{footnote}}
\setcounter{footnote}{0}
\vspace{6ex}

%--------------------<<   abstract   >>--------------------
\begin{abstract}
We show that the unification of the doublet Higgs in the standard model (SM) 
and the Higgs to break the grand unified theory (GUT) group stabilizes
the sliding singlet mechanism which can solve the doublet-triplet~(DT)
splitting problem. And we generalize this attractive mechanism
to apply it to many unified scenarios. In this paper, we try to build
various concrete $E_6$ unified models by using the generalized
sliding singlet mechanism.

\end{abstract}

\end{titlepage}

%--------------------<<   section    >>--------------------

\section{Introduction}

The well-known success of the gauge coupling unification
in the minimal supersymmetric standard model~(MSSM) 
likely supports the attractive idea of supersymmetric 
grand unified theory~(SUSY-GUT).
On the other hand, we know there are some obstacles 
in constructing a realistic SUSY-GUT. 
One of the biggest problems is the so-called DT splitting 
problem.
Generically in SUSY-GUTs, there are color triplet 
partners of the MSSM Higgs, and the nucleon decay via dimension
five operators becomes too rapid.
In order to suppress this proton decay, the color triplet partners 
must have very large mass $(\gg\Mu\sim10^{16}\GeV)$, 
in contrast to the doublet Higgs whose mass has to be of order 
the weak scale $\MW\sim10^2\GeV$.
Some ideas to solve this problem have been proposed : 
the sliding-singlet mechanism\cite{SS,Sen,su6,su6su2}, 
the missing partner mechanism\cite{missing,flippedsu5}, 
the Dimopoulos-Wilczek (DW) mechanism\cite{DW}, the GIFT 
mechanism\cite{psudeNG}, and via orbifold 
boundary condition\cite{5dDTS}.

Among these ideas, the first mechanism is the smartest solution 
which realizes the DT splitting dynamically. 
Although it was shown that the originally proposed $SU(5)$ model 
cannot act effectively if SUSY breaking effect is 
considered\cite{SBeffect}, some authors have proposed $SU(6)$ extensions 
in which this mechanism acts without destabilization 
due to SUSY breaking\cite{Sen,su6,su6su2}. 
In this paper, we abstract the essence of this sliding singlet 
mechanism in $SU(6)$ models and generalize it to apply to many other 
unified theories.
Actually in $E_6$ unification it is found that for many directions of 
VEV of the adjoint Higgs this mechanism may act.
Corresponding to these breaking patterns, we construct some 
\E6 Higgs sectors in which the DT splitting problem is 
indeed solved through this mechanism. 
Several concrete models are propose in the context of the 
SUSY-GUT in which an anomalous $U(1)_A$ gauge symmetry\cite{U(1)},
whose anomaly is cancelled by the Green-Schwarz mechanism\cite{GS},
plays an important 
role\cite{NGCU,BM,horiz,reduced,maekawa,maekawa2,flippedso10} in
solving various problems in SUSY scenario.
And we examine whether the already proposed 
realistic quark and lepton sector\cite{BM} is compatible 
with such a Higgs sector or not.
Note that this \E6 group is interesting as a unified group, 
in the sense that the SUSY flavor problem can be solved 
in \E6 SUSY-GUT with anomalous $U(1)_A$ and non-abelian horizontal 
gauge symmetry\cite{horiz}. 

In section 2, we briefly review the sliding singlet mechanism in 
the context of $SU(5)$ and $SU(6)$.
In section 3, we generalize this mechanism to the general gauge 
group. 
In section 4, we construct some concrete Higgs sectors.

\section{The Sliding Singlet Mechanism}

In this section, We review the present status of 
the sliding singlet mechanism. 
For this purpose, we sometimes omit details, which are described 
in each references.

\subsection{$SU(5)$}

The sliding singlet mechanism was originally proposed in the context 
of $SU(5)$\cite{SS}, in which the following terms are allowed in the 
superpotential;
\begin{equation}
W_{\mbox{ss}} = \bar{H} ( A + Z ) H.
\label{su5ss}
\end{equation}
Here, the adjoint Higgs $A(\bf{24})$ is assumed 
to have the VEV 
$\VEV A={\rm diag}(2,2,2,-3,-3)v$ which breaks 
$SU(5)$ into \Gsm\ ($G$\sub{SM}), and the (anti)fundamental Higgs 
$H({\bf5})$ and $\bar H({\bf\bar5})$ 
contain the MSSM doublet Higgs, $H_u$ and $H_d$, respectively.
Since the doublet Higgs have non-vanishing VEVs $\VEV{H_u}$ and 
$\VEV{H_d}$ to break 
$SU(2)$\sub L$\times U(1)$\sub Y into $U(1)$\sub{EM}, 
the minimization of the potential,
\bequ
V_{{\mbox{\scriptsize{SUSY}}}}=\abs{F_H}^2+\abs{F_{\bar H}}^2=
\Ls\abs{\VEV{\bar H}}^2+\abs{\VEV H}^2\Rs\abs{-3v+\VEV Z}^2,
\label{potential}
\eequ
leads to the vanishing doublet Higgs mass  
$\mu=(\VEV A+\VEV Z)_{\bf2}=-3v+\VEV Z=0$
by {\it sliding} the VEV of the singlet Higgs 
$Z({\bf1})$\footnote{
Here, the contributions to the potential from 
$F_A$ and $F_Z$ are neglected, because they are of order $(\VEV{\bar HH})^2$.
In this sense, the doublet Higgs mass $\mu$ does not vanish exactly but may 
become
of order SUSY breaking scale $\MSB$.
}.
For these VEVs, $\VEV A+\VEV Z=\diag(5,5,5,0,0)v$, the color 
triplet partners of doublet Higgs have a large mass 
$5v\sim 10^{16}\GeV$.

Unfortunately, it is known that if SUSY breaking is taken into account,
this DT splitting is failed. For example, 
the soft SUSY breaking mass term $\s m^2\abs Z^2$ 
$(\s m\sim\MSB)$ 
shifts the VEV $\VEV Z$ by an amount of 
$\delta\VEV Z\sim\frac{\s m^2v}{{\VEV H}^2+\s m^2}\sim\Mu$ to minimize
the potential.
Thus the doublet-triplet splitting is spoiled by SUSY breaking effect
in this mechanism.
\footnote{
The soft term $\s mZF_Z$ also destabilizes the sliding singlet
mechanism because this term alters the contribution of 
$F_Z$ to the scalar potential as $\abs{\bar HH+\s mZ}^2$, that
is the order of $\MSB^2\Mu^2(\gg\MSB^4)$ if $\VEV H$ is 
order of the weak scale.
Such a term is induced by loop effects through the coupling between
$Z$ and the color triplet Higgs. Therefore, even if the terms
$\s mZF_Z$ and $\s m^2|Z|^2$ are absent at the tree level, this problem
cannot be avoided.
}
This is caused by the fact that 
the terms $|F_H|^2+|F_{\bar H}|^2$ give only a mass of 
order ${\VEV H}$ to $Z$,
which are the same order as (or smaller than) 
the SUSY breaking contribution.
Since this mass parameterizes the stability of $\VEV Z$ 
against other contributions to the potential, \eg\ 
SUSY breaking effects
$\s m^2 |Z|^2$, 
soft terms of order $\MSB$ easily shift the VEV from that 
in the SUSY limit by a large amount.

It is obvious that this problem can be solved if we take
the large VEVs $\VEV H$ and $\VEV {\bar H}$ (larger than
$\sqrt{\MSB\Mu}$) which give larger mass to
$Z$ and stabilize the VEV of $Z$ against SUSY breaking
effects. 
Of course, it is not consistent with the experiments to
take such large VEVs for SM doublet Higgs.
But for other Higgs, for example, that breaks
a larger gauge group into the SM gauge group, 
we can take such large VEV. This is an essential idea of
Sen\cite{Sen}.

\subsection{$SU(6)$}
\label{su6}

$SU(6)$ is the simplest candidate for the above purpose, and some 
authors examine the possibility\cite{Sen,su6}.
The relevant part of the superpotential is written in a similar form 
as (\ref{su5ss}), where $A({\bf35})$ has a VEV 
$\VEV A={\rm diag}(1,1,1,-1,-1,-1)v$ which breaks 
$SU(6)$ into $SU(3)$\sub C$\times SU(3)$\sub L$\times U(1)$ and 
$H({\bf6})$ ($\bar H({\bf\bar6})$) denotes (anti)fundamental Higgs 
which has a VEV in $SU(5)$ singlet component. 
Note that the simplest extension, 
in which one pair of (anti)fundamental Higgs is introduced, 
cannot act effectively. 
This is because 
\bit
 \item the $F$-term of $Z$ gives a contribution of \order{\Mu^4} 
       to the scalar potential.
 \item the term $\bar HAH$ gives a contribution to the $F$-flatness 
       condition of $A$ which destabilizes the required form of VEV 
       of $A$, if $\VEV A$ is determined from $F_A$.
 \item this term also gives mass terms of {\bf5} and {$\bf\bar5$} 
       of $SU(5)$, $\VEV{\bar H}AH$ and $\bar HA\VEV H$.       
 \item one pair of doublet of (anti)fundamental Higgs 
       %with non-vanishing VEV will be 
       is the Nambu-Goldstone (NG) mode 
       and unphysical.
\eit
The first two and the last one are resolved if one of 
the pair of (anti)fundamental Higgs has vanishing VEV.
Thus, (\ref{su5ss}) is altered as
\begin{equation}
W_{\mbox{ss}} = \bar{H'} ( A + Z ) H + \bar{H} ( A + Z ) H', 
\label{su6ss}
\end{equation}
where primed fields have vanishing VEVs.
Because of the third reason, at least one more pair of 
(anti)fundamental Higgs is required. 

For example in Ref.\cite{su6}, four pairs are introduced and 
the relevant part of the superpotential is given as 
\beqn
 W & = & W(A) + W(\bar{H}_i, H_i) \nn\\
   & & + \sum_{i = 1,2} a_i\bar{H'}_i( A + Z_i) H_i
       + \sum_{i = 1,2} \bar a_i\bar{H}_i (A + \bar{Z}_i) H'_i,
\eeqn
where $a_i$ and $\bar a_i$ are coupling constants and 
$W(A)$ and $W(\bar{H}_i, H_i)$ are some sets of terms 
which give the desired VEV (as one of discrete vacua) to 
$A$ and $\bar{H}_i, H_i$ respectively.
This gives following mass matrix of ${\bf5}\times{\bf\bar5}$ 
of $SU(5)$:
\bequ
  M_I = \bordermatrix{
          & I_A & I_{\bar H'_1} & I_{\bar H'_2} 
                & I_{\bar H_1} & I_{\bar H_2}  \cr
       \bar I_A & M_I & \bar a_1\VEV{\bar H_1} 
                & \bar a_2\VEV{\bar H_2}& 0 & 0  \cr
  \bar I_{H'_1} & a_1\VEV{H_1}& 0 & 0 & 2\alpha_I a_1v & 0  \cr
  \bar I_{H'_2} & a_2\VEV{H_2}& 0 & 0 & 0 & 2\alpha_I a_2v  \cr
   \bar I_{H_1} & 0 & 2\alpha_I \bar a_1v & 0 &c\abs{\VEV{H_2}}^2 
                & -c\VEV{H_1 H_2^*}  \cr
   \bar I_{H_2} & 0 & 0 & 2\alpha_I \bar a_2v & -c\VEV{H_1^* H_2} 
                & c\abs{\VEV{H_1}}^2 \cr
            }.  
\eequ
Here $\alpha_{\bf2}=0$ and $\alpha_{\bf3}=1$, which are realized
by the sliding singlet mechanism,
and $M_{\bf3}=0$ because ${\bf 3}_A$ and ${\bf \bar 3}_A$ are NG
modes by breaking $SU(6)\rightarrow SU(3)_C\times SU(3)_L\times U(1)$.
$M_{\bf2}$ and $c$ are determined 
from $W(A)$ and $W(\bar{H}_i, H_i)$, respectively. From 
this mass matrix, it can be found that there are 
two massless modes for $I={\bf2}$ and 
one for $I={\bf3}$. 
Since one pair of {\bf5} and {$\bf\bar5$} of $SU(5)$ are absorbed by 
the Higgs mechanism, only one pair of doublets remains massless 
and therefore the DT splitting is realized.
In this model, the massless modes come from a linear combination 
of the primed fields. 

Note that, due to the large VEVs of $H_i$ and $\bar H_i$, 
this hierarchy is stable against the SUSY breaking corrections, 
which means that all the elements of the mass matrix have corrections 
due to SUSY breaking at most \order{\MSB}.
\\

Another example was proposed in Ref.\cite{su6su2} in 
the context of $SU(6)\times SU(2)$.
The relevant part is similar as previous model except the absence 
of the indices $i$ of the singlet Higgs. 
The indices of the (anti)fundamental Higgs are understood as 
those of the symmetry $SU(2)$.
This symmetry guarantees that the doublet components of $H_2$ and 
$\bar H_2$ are massless even if they do not have non-vanishing VEVs.
This means that the doublets are not NG modes and 
therefore physical modes.
In this model, the SM doublet Higgs comes from the unprimed fields 
which have non-vanishing VEVs, in contrast to the previous model.%
\footnote{
In order to give masses to the primed fields, 
some additional terms, \eg\ $\bar H'_i H'_i$, are needed. 
}

The author of Ref.\cite{su6su2} mentions that this is because the 
symmetry $SU(2)$ relates the doublets to NG modes 
in $H_1$ and $\bar H_1$. 
To be more precise, the doublets and the NG modes belong to 
a single multiplet of the $SU(2)$ symmetry to which the mass parameter 
$(\VEV A+\VEV Z)$ respect.
However, in the spirit of the sliding singlet, 
it may be more appropriate to say that the mass parameter $(\VEV A+\VEV Z)$ 
gives the same value for the doublets as for the $SU(5)$ singlet 
components of $H_1$ and $\bar H_1$, which must vanish due to the 
non-vanishing VEVs.
This observation makes it possible to apply the sliding singlet 
mechanism in more general case.

\section{Generalization}

Now, we examine how we can generalize the sliding singlet mechanism. 

The essential idea of the sliding singlet mechanism is following.
\bit
 \item If the mass parameter of a certain component is guaranteed to be
       the same value 
       as that of the other component which has a non-vanishing VEV and 
       the later vanishes dynamically due to the VEV, 
       the former also vanishes. 
       And if the non-vanishing VEVs are sufficiently large, 
       the mass hierarchy is stable against possible 
       SUSY breaking effects.
\eit
It is the case that a doublet component and a singlet component 
with non-vanishing VEV belong to a single multiplet of the symmetry 
to which the mass terms respect, 
\eg\ $SU(3)$\sub C$\times SU(3)$\sub L$\times U(1)$ 
for the previous $SU(6)$ example in Ref.\cite{su6}. 
In this case, the DT splitting problem can be solved. 

Moreover, if the mass parameter depends only on the VEVs of adjoint Higgs 
and singlet Higgs, the above condition for the sliding singlet 
mechanism to act can be easily examined.
This is because the mass parameter for each component is 
determined by each quantum number of $U(1)$ which are fixed by
the non-vanishing VEV of the adjoint Higgs.
Therefore, if the charge of the doublet Higgs component is the same as
that of the singlet component which has non-vanishing VEV, then
the massless doublet Higgs can be realized by the sliding singlet 
mechanism.
This perspective holds for any gauge group, even if the mass term 
involves non-renormalizable terms. 

It is obviously important to know the charges for the SM singlet and 
the doublet Higgs under the $U(1)$ generator which are fixed 
by the non-vanishing VEV of the adjoint Higgs. 
If a GUT group $G$ includes 
\Gsm$\times U(1)^{r-4}$ as a subgroup, where $r$ is the rank of $G$, 
the $U(1)$ generator must be a linear combination of $r-3$ $U(1)$
generators. Therefore, it is helpful to know the charges of these
$U(1)$ generators in order to classify the models in which sliding
singlet mechanism acts effectively. We present the charges for 
$SU(6)$ unification group in Table \ref{qsu6}, and 
for $E_6$ unification group in Table \ref{qe6}\footnote{
We denote each representation of $G$\sub{SM}, 
$({\bf 3,2})_{\frac{1}{6}}$,
$({\bf \cc 3,1})_{-\frac{2}{3}}$, $({\bf \cc 3,1})_{\frac{1}{3}}$,
$({\bf 1,2})_{-\frac{1}{2}}$, $({\bf 1,1})_1$, $({\bf 1,1})_0$,
$({\bf 3,2})_{-\frac{5}{6}}$, $({\bf 8,1})_0$, and $({\bf 1,3})_0$
as $Q$, $U$, $D$, $L$, $E$, $N$, $X$, $G$, and $W$, respectively, 
and the conjugate representation of a representation $R$ by $\bar R$. 
}.

\begin{table}[th]
\bctr
\begin{tabular}{|c||c|c||c|c|}\hline
 $SU(6)$ & \multicolumn{2}{c||}{\bf6} & \multicolumn{2}{c|}{\bf35}
 \\ \hline \hsp0\vsp{-0.45}&&&& \\ 
 $SU(5)$ & {\bf5} & {\bf1} 
         & ${\bf24}$ & ${\bf\bar5}$ 
 \\ \hline \hsp0\vsp{-0.45}&&&& \\ 
 $G$\sub{SM} & $\bar D, \bar L$ & $N$
             & $X$ & $D,L$ 
 \\ \hline\hline
 $V_6$ & 1 & $-5$ & 0 & $-6$ 
 \\ \hline
 $6Y$ & $-2,3$ & $0$ 
      & $-5$ & $2,-3$ 
 \\ \hline \hline
 $V_6-12Y$ & $5,-5$ & $-5$ 
           & 10 & $-10,0$
 \\ \hline
\end{tabular}
\ectr
\caption{
The charges of the relevant $U(1)$ for $SU(6)$ model.
Here, $V_6$ is defined by the relation $SU(6)\supset SU(5)\times U(1)_{V_6}$
and we omit representations $G$ and $W$ which have trivial charges 
and ${\bf 5}$ in ${\bf 35}$ which is a conjugated field of ${\bf\bar 5}$ 
in ${\bf 35}$.
}
\label{qsu6}
\end{table} From Table \ref{qsu6}, we can see that there is one possible 
breaking pattern, $V_6-12Y$ direction, 
\ie\ $\VEV A \propto \diag(1,1,1,-1,-1,-1)$, for the sliding singlet 
mechanism to act in the mass terms of ${\bf 6}$ and ${\bf \bar 6}$.
In other words, only for the $U(1)$ generated by the generator of this 
direction, the singlet and doublet of $G$\sub{SM} have the same charge.

\begin{table}[th]
\bctr
\begin{tabular}{|c||c|c|c|c|c|c||c|c|c|c|c|}\hline
 \E6 & \multicolumn{6}{c||}{\bf27} & \multicolumn{5}{c|}{\bf78}
 \\ \hline
 $\!SO(10)\!$ & \multicolumn{3}{c|}{\bf16} & \multicolumn{2}{c|}{\bf10} 
          & {\bf1}
          & \multicolumn{2}{c|}{\bf45} & \multicolumn{3}{c|}{\bf16}
 \\ \hline \hsp0\vsp{-0.45}&&&&&&&&&&& \\ 
 $SU(5)$ & {\bf10} & ${\bf\bar5}$ & {\bf1} & {\bf5} & ${\bf\bar5}$ 
         & {\bf1}
         & $\!{\bf24}\!$ & {\bf10} & {\bf10} & ${\bf\bar5}$ & {\bf1}
 \\ \hline \hsp0\vsp{-0.45}&&&&&&&&&&& \\ 
 $G$\sub{SM} & $\!Q,U,E\!$ & $\!D,L\!$ & $\!N\!$ 
             & $\!\bar D, \bar L\!$ & $\!D,L\!$ 
             & $\!\!N\!\!\!$
             & $X$ & $\!Q,U,E\!$ 
             & $\!Q,U,E\!$ & $\!D,L\!$ & $\!N\!$ 
 \\ \hline\hline
 $V'$ & \multicolumn{3}{c|}{1} & \multicolumn{2}{c|}{$-2$} 
      & 4
      & \multicolumn{2}{c|}{0} & \multicolumn{3}{c|}{$-3$}
 \\ \hline
 $V$ & 1 & $-3$ & 5 & $-2$ & 2 
     & 0
     & 0 & $-4$ & 1 & $-3$ & 5 
 \\ \hline
 $6Y$ & $\!1,\!-4,6\!$ & $\!2,\!-3\!$ & $0$ & $\!\!-2,3\!$ & $\!2,\!-3\!$ 
      & $0$ 
      & $\!\!-5\!$ & $\!1,\!-4,6\!$ & $\!1,\!-4,6\!$ & $\!2,\!-3\!$ & $0$ 
 \\ \hline\hline \hsp0\vsp{-0.45}&&&&&&&&&&& \\ 
 $\!\!\!$ I:$\frac{(0,3,12)}{5}$ 
      & $\!1,\!-1,3\!$ & $\!\!-1,\!-3\!$ & $3$ & $\!\!-2,0\!$ & 2, 0 
      & 0
      & $\!\!-2\!$ & $\!\!-2,\!-4,0\!$ 
       & $\!1,\!-1,3\!$ & $\!\!-1,\!-3\!$ & \ 3 
 \vsp{-0.45}\\&&&&&&&&&&& \\ \hline \hsp0\vsp{-0.45}&&&&&&&&&&& \\ 
 $\!\!\!$ II:$\frac{(5,3,-48)}{20}\!$ 
      & $\!0,2,\!-2\!$ & $\!-1,1$ & 1 & $\!0,\!-2\!$ & $-1,1$ 
      & 1 
      & $\!2\!$ & $\!\!-1,1,\!-3\!$ & $\!\!-1,1,\!-3\!$ & $-2,0$ & \ 0 
 \vsp{-0.45}\\&&&&&&&&&&& \\ \hline \hsp0\vsp{-0.45}&&&&&&&&&&& \\ 
 $\!\!\!$ III:$\frac{(5,-9,24)}{20}\!$ 
      & $\!0,\!-1,1\!$ & 2, 1 & $\!\!-2\!$ & 0, 1 & $\!\!-1,\!-2\!\!$ 
      & 1 
      & $\!\!-1\!$ & 2, 1, 3 & $\!\!-1,\!-2,0\!$ & 1, 0 
       & $\!\!\!-3\!\!\!\!$ 
 \vsp{-0.45}\\&&&&&&&&&&& \\ \hline \hsp0\vsp{-0.45}&&&&&&&&&&& \\ 
 $\!\!\!$ IV:$\frac{(5,3,72)}{20}\!$ 
      & $\!1,\!-2,4\!$ & $1,-2$ & 1 & $\!\!-2,1\!\!$ & $1,\!-2$ 
      & 1 
      & $\!\!-3\!$ & $0,-3,3$ & $0,-3,3$ & $0,\!-3$ & \ 0 
 \vsp{-0.45}\\&&&&&&&&&&& \\ \hline
\end{tabular}
\ectr
\caption{
The charges of the relevant $U(1)$ for \E6 model.
Here, we omit representations ${\bf \overline{10}}$ in ${\bf 45}$ which
is a conjugated field of ${\bf 10}$ in ${\bf 45}$, and the fields omitted
in Table 1.
In the first column of the last four rows, $\frac{(a,b,c)}{d}$ 
denotes the linear combination 
$\frac{a}{d}V'+\frac{b}{d}V+\frac{c}{d}Y$,%
where $V'$ and $V$ are defined by  the relations
$E_6\supset SO(10)\times U(1)_{V'}\supset 
SU(5)\times U(1)_V\times U(1)_{V'}$.
}
\label{qe6}
\end{table}
On the other hand, as seen in Table \ref{qe6}, there are many 
possible breaking patterns for the generalized sliding singlet mechanism
in $E_6$ models.
Actually there are infinite possibilities because there is only one relation,
namely, the charge of a singlet is equal to that of a doublet, and
three independent $U(1)$'s in $E_6$ unified models.
If we impose two relations, the $U(1)$ can be fixed except for the 
normalization. For example, we require that the charge of a
singlet is equal to the charges of two doublets.
Because there are two singlets and three doublets of SM gauge group 
in a single field ${\bf 27}$ of $E_6$, there are several solutions
as in Table \ref{qe6}. Essentially there are three solutions ( Model 
I, II, and III in Table \ref{qe6}).
The Model I gives the DW type of VEV which breaks $E_6$ into 
$SU(3)_C\times SU(2)_L\times SU(2)_R\times U(1)_{B-L}\times U(1)_{V'}$, 
and the models in Ref.\cite{reduced} are classified as this case. 
The breaking pattern of Model II is that \E6 is broken into 
$SU(3)$\sub C$\times SU(3)$\sub L$\times U(1)\times SU(2)$\sub E, 
where $SU(2)$\sub E rotates two ${\bf\bar5}$s and {\bf1}s of $SU(5)$ 
in {\bf27} of \E6 as the doublets. 
This produces similar models as in Ref.\cite{su6su2} if $SU(2)$\sub E 
is identified as the symmetry $SU(2)$ in Ref.\cite{su6su2}.
The Model III preserves 
$SU(3)$\sub C$\times SU(3)$\sub L$\times SU(2)$\sub {RE}$\times U(1)$, 
where $SU(2)$\sub {RE} denotes the $SU(2)$ sub-group of 
$SU(3)$\sub R under which the component of {\bf3} which belongs to 
the doublet of both $SU(2)$\sub R and $SU(2)$\sub E is singlet.
If we require that $SU(2)_E$ symmetry is not broken by the VEV of the
adjoint Higgs  and the charge of a singlet is equal to the charge of a
doublet, then in addition to Model II, Model IV is satisfied with the 
requirements. 
In the Model IV, \E6 is broken into $SO(10)$\sub{F}$\times U(1)$.%

\vsp{0}From the above observation, models in which the DT 
splitting is realized through the sliding singlet mechanism may 
be constructed for each breaking pattern, 
because some doublet components $L$ and/or $\bar L$ in {\bf27} 
have the same charge as some singlet components $N$. 
However, as illustrated in the previous section, this is not a sufficient
condition. Actually we have to take care of the following points 
in constructing models:
\bit
 \item the massless mode may be an unphysical NG mode.
 \item other part of superpotential may give a large mass to the 
       would-be massless mode.
 \item an effort to avoid the above two obstructions may yield 
       unwanted additional massless modes.
\eit
In the next section, we try to construct several models in which
DT splitting is realized.

\section{Models}

Here we construct concrete models corresponding to 
the breaking patterns listed in Table \ref{qe6} 
in the context of anomalous $U(1)_A$ gauge symmetry. 
Since, in GUTs with anomalous $U(1)_A$ gauge symmetry with generic 
interaction, positively charged fields have vanishing VEVs% 
\cite{BM,maekawa,maekawa2}, 
they naturally play the role of the primed fields.

\subsection{Model I: \Ga$\times U(1)$\sub{V'}}
\label{e6dw}

Recently we proposed a \E6 unified scenario in which the DT splitting 
problem is solved by the DW mechanism\cite{reduced}.
In Ref.\cite{reduced}, we emphasized that the sliding singlet
mechanism is also workable in the model but focused on the DW mechanism.
However, according to the new perspective proposed in the previous section, 
the DT splitting in the model can be understood only by using the generalized
sliding singlet mechanism.

We consider the Higgs sector defined by Table \ref{model}, 
where lowercase letters denotes the anomalous $U(1)_A$ charge, 
and $\Theta$ is the Froggatt-Nielsen field\cite{FN}. 
\begin{table}[th]
\begin{center}
\begin{tabular}{|c|c|} 
\hline 
{\bf 78}          &   $A(a=-1,+)$\ $A'(a'=5,+)$       \\
{\bf 27}          &   $\Phi(\phi=-3,+)$\  $C'(c'=6,-)$  \\
${\bf \overline{27}}$ &$\bar \Phi'(\bar \phi'=5,+)$ 
                     \ $\bar C(\bar c=0,-)$  \\
{\bf 1}           &   $\Theta(\theta=-1,+)$\ $Z_i(z_i=-1,+)$   \\
\hline
\end{tabular}
\end{center}
\caption{
Typical values of anomalous $U(1)_A$ charges. 
Here, $\pm$ denotes the quantum number of an additional $Z_2$ parity.
}
\label{model}
\end{table}From this table, 
we can lead the relevant superpotential to determine 
VEVs%
\footnote{
See Ref.\cite{reduced} and its references% 
\cite{BM,maekawa,maekawa2}. 
}
:
\begin{equation}
  W=W_{A^\prime} + W_{\bar\Phi^\prime}+W_{C^\prime},
\label{W}
\end{equation}
where 
\beqn
  W_{A'} &=& A'(A+A^3+A^4+A^5) 
\label{Wa} \\
  W_{\bar\Phi'} &=& \bar\Phi'(1+A+Z_i+A^2+AZ_i+Z_i^2)\Phi 
\label{Wphi}\\
  W_{C'} &=& \bar C(1+A+Z_i+\cdots+(\bar C\Phi)^2)C'. 
\label{Wc}
\eeqn
Combining with the $D$-flatness conditions, we can see the following 
VEVs 
\beqn
  &\VEV{{\bf\cc{16}}_{\bar C}} 
    \sim \VEV{{\bf16}_A}
    \sim \VEV{{\bf1}_\Phi}
    \sim \lambda^{-a}\lambda^r & \label{VEV1}\\
  &\VEV{{\bf{1}}_{\bar C}}
    \sim \VEV{{\bf16}_\Phi}
    \sim \lambda^{-a}\lambda^{2r} & \label{VEV2}\\
  &\VEV{{\bf45}_A} \sim \lambda^{-a}& \label{VEV3}\\
  &\mbox{Other VEVs} =0&
\eeqn
are indeed one of vacua of this model. 
Here, $r=\frac{2a-\phi-\bar c}{3}$, $\lambda$ is the ratio of 
the VEV of $\Theta$ to the cutoff scale $\Lambda$, and
the diagonal part of the VEV of $A$ points to the direction $V+4Y$, 
$\VEV{{\bf 45}_A}=\tau_2\times{\rm diag} (1,1,1,0,0)v$%
~(DW form), 
which breaks \E6 into \Ga$\hsp0$ $\times U(1)_{V'}(\equiv G_A)$.
Here, $\tau_2$ denotes the second Pauli matrix.

For these VEVs, the sliding singlet mechanism acts in $W_{\bar\phi'}$ 
as mentioned in the previous section. 
One of the $F$-flatness conditions from $W_{\bar\phi'}$, 
\begin{eqnarray}
F_{{\bf 1}_{\bar \Phi'}}&=& (1+A+Z_i+A^2+AZ_i+Z_i^2)_{\bf1}{\bf1}_\Phi=0, 
\label{Fphi1} 
\end{eqnarray}
makes the mass terms of doublet components in 
${\bf10}_{\bar\Phi'}\times{\bf10}_\Phi$ of $SO(10)$ vanishing. 
This is because they have the same charges as ${\bf1}_\Phi$ 
(which happen to be zero%
\footnote{
Because of the zero, the DW mechanism can act in $SO(10)$ model 
if only ${\bf 10\cdot45\cdot10}$ contributes the doublet mass terms.
In order to forbid the dangerous terms, \eg\ ${\bf 10\cdot10}$, 
the additional symmetry, \eg\ $Z_2$ parity, is required 
in $SO(10)$ model. 
On the other hand, in \E6 model, 
the prohibition can be realized dynamically.
}) 
of\ $U(1)$ which determined by $\VEV A$, \ie\ $U(1)_{B-L}$, 
and therefore the mass parameters become the same value as for 
${\bf1}_{\Phi'}\times{\bf1}_\Phi$, 
which vanishes due to Eq.(\ref{Fphi1}).
Note that because ${\bf {16}}_A$ has non-vanishing VEV,
we have to examine carefully the other $F$-flatness conditions of 
${\bf \overline{16}}_{\bar\Phi'}$ in which the VEV $\VEV{{\bf {16}}_A}$
appears. As we pointed out in Ref.~\cite{reduced}, because of the $E_6$ group
theoretical reason and the $F$-flatness condition (\ref{Fphi1}),
the $F$-term is factorized
as 
\begin{equation}
F_{{\bf \overline{16}}_{\bar \Phi'}}=(1+Z_i+A)({\bf 45}_A{\bf 16}_\Phi
                                     +{\bf 16}_A{\bf 1}_\Phi)=0
\label{F16}
\end{equation}
in this vacuum. 
Therefore, in this model, the sliding singlet mechanism acts for the singlets
in ${\bf 16}_\Phi$ and ${\bf 16}_A$, namely, the factor $(1+Z_i+A)$ in Eq. 
(\ref{F16}) vanishes by sliding the singlet
VEV $\VEV{Z_i}$. Actually, in the model in Table {\ref{model}}, two $E$ fields
of ${\bf 16}_{\Phi}$ and ${\bf 16}_A$, which have the same $U(1)$ charges as
$N$ in ${\bf 16}_\Phi$ and ${\bf 16}_A$, respectively, become 
massless, though they are absorbed
by the Higgs mechanism. 
In order to confirm that only one pair of doublet is massless, 
we have to check all mass matrices explicitly. 
Straightforward calculation shows it,
but we would like to skip it here.
And we just mension that we have constructed a model which solves the DT 
splitting 
problem by using the generalized sliding singlet mechanism.
We stress that, in this model, the massless Higgs doublets come from a 
single multiplet $\Phi({\bf 27})$.
To be more precise, they come from ${\bf10}_\Phi$ which is not 
related to any NG modes by the remaining symmetry $G_A$. 
This situation is quite different from the $SU(6)$ cases in which
by the sliding singlet mechanism a pair of doublets becomes massless
but is absorbed by the Higgs mechanism. 
Note that in mass terms generated from $W_{C'}$,  
generally the sliding singlet mechanism does not act, since the VEV 
of $\bar C\Phi$ respects only $G$\sub{SM}.

And, as mentioned in Ref.\cite{reduced}, this Higgs sector is 
compatible with the matter sector proposed in Ref.\cite{BM}.
This is because the main modes of doublet Higgs come from a fundamental
representation field $\Phi({\bf 27})$ and not from an anti-fundamental 
fields $\bar C({\bf \overline{27}})$. This fact results in comparatively
large Yukawa couplings which are important to realize large top Yukawa 
coupling and to avoid too small $\tan \beta$.

\subsection{Model II:
$SU(3)$\sub C$\times SU(3)$\sub L$\times U(1)\times SU(2)$\sub {E}}
\label{e6de}

This breaking pattern is similar as that of the SU(6) models 
in \S\ref{su6}, and therefore the reason that the massless doublet 
Higgs appear can be understood in the similar way.

We consider the Higgs sector defined by Table \ref{model}.
Though we adopt the same anomalous $U(1)_A$ charges as in Table \ref{model},
we take different number of the singlet Higgs fields.
The difference between the previous vacuum and this vacuum is essentially
that in this vacuum, the diagonal part of the VEV of $A$, 
$\VEV{({\bf45+1})_A}$,  points to the 
direction $5V'+3V-48Y$, which breaks \E6 into 
$SU(3)$\sub C$\times SU(3)$\sub L$\times U(1)\times SU(2)$\sub {E}. 
For this vacuum, possible candidates for the NG mode in the 
representation $L$ ($\bar L$) are from the fields $\Phi$ 
($\bar C$) and not from $A$.
Therefore, at least one of the two $L$ of $\Phi$ is the NG mode, namely becomes
massless. On the other hand, the two $L$ of $\Phi$ are doublet under the
the symmetry $SU(2)$\sub {E}, that leads to the both of the $L$ must be 
massless. One of the two $L$ is absorbed by the Higgs mechanism, but the other
$L$ becomes a physical massless mode
\footnote{
On the other hand, because the mass term of $\bar C$ does not respect 
$SU(2)$\sub {E} due to the term $\bar C(\bar C\Phi)^2C'$, 
one linear combination of the two $\bar L$ of $\bar C$ have a 
non-vanishing mass parameter.
}.

On the other hand, according to the new perspective proposed 
in the previous section, this can be understood in the different way.
Namely, the vanishing mass term is caused by the $F$-flatness 
condition (\ref{Fphi1}).
This condition makes the two mass terms for  
${\bar L}_{{\bf 10}_{\bar\Phi'}}\times L_{{\bf 10}_\Phi}$ and
${\bar L}_{{\bf \overline{16}}_{\bar\Phi'}}\times L_{{\bf 16}_\Phi}$
  also vanishing, 
due to their same charges as  ${\bf1}_\Phi$.
Because ${\bf 16}_A$ has non-vanishing VEV, we have to take care of
the $F$-flatness condition of ${\bf \overline{16}}_{\bar\Phi'}$,
\begin{equation}
F_{{\bf \overline{16}}_{\bar \Phi'}}=(1+A+Z_i+A^2+AZ_i+Z_i^2){\bf 16}_\Phi
                                     +(1+Z_i+A){\bf 16}_A{\bf 1}_\Phi=0.
\label{Fphi16}
\end{equation}
Note that in this breaking pattern, this $F$-term is not factorized.
(This is because the charge of ${\bf 1}_\Phi$ does not vanish in this vacuum,
but that vanishes in the previous vacuum.) 
The first term in Eq. (\ref{Fphi16}) vanishes by the sliding singlet mechanism
because ${\bf 16}_\Phi$ has the same $U(1)$ charge as ${\bf 1}_\Phi$.
Therefore the second term must vanish by itself and the VEV of a singlet field
again slides to satisfy this relation. 
The coefficient $(1+Z_i+A)$ of the second term is equivalent to that of
the term 
${\bar L}_{{\bf \overline{10}}_{\bar \Phi'}}\VEV{{\bf 16}_A}L_{{\bf 16}_\Phi}$,
because ${\bf 16}_A$ has vanishing $U(1)$ charge and $L_{{\bf 16}_\Phi}$
has the same $U(1)$ charge as ${\bf 1}_\Phi$. Because of these effects,
the two $L_\Phi$ become massless.
One of the two $L_\Phi$ are absorbed by the Higgs mechanism and the other 
becomes a physical massless doublet Higgs.
Because the other part does not contribute to the mass of $\Phi$ and 
it can be checked straightforwardly that there are no additional 
massless modes, the DT splitting is realized.

In this model, the massless $L$ comes from $\Phi$ and 
the massless $\bar L$ comes from a certain linear combination, 
the main mode of which comes from a primed field. This is because positively
charged fields (primed fields) have smaller couplings and therefore smaller
masses than negatively charged fields (unprimed fields). 
Unfortunately we do not know a realistic quark and lepton sector 
compatible with this model, in contrast to the model in the previous 
subsection. The biggest difference is that the main component of 
the ${\bar L}$ Higgs comes from the primed fields 
and not from unprimed fields in this model. And therefore,
the top Yukawa coupling is suppressed because the Higgs $\bar L$ has
positive charge. It is the essential reason for this 
suppression that the sliding singlet mechanism does not act for
${\bar L}$ in ${\bf 27}$. To avoid this situation, it is important
to take the same $U(1)$ charge of ${\bar L}$ in ${\bf 27}$ as
that of ${\bf 1}$ in ${\bf 27}$. 

In the following, we examine models in which ${\bar L}$ in ${\bf 27}$ 
has the same charge as ${\bf 1}$ in ${\bf 27}$.

\subsection
{Model III:
$SU(3)$\sub C$\times SU(3)$\sub L$\times SU(2)$\sub {RE}$\times U(1)$}

This is the last breaking pattern in which the sliding singlet mechanism
acts for two doublet components in ${\bf 27}$. Because one of them is 
${\bar L}$, large top Yukawa coupling can be expected to be realized.
Here we consider the same Higgs sector as in \S\ref{e6dw} and \S\ref{e6de} 
defined by Table \ref{model} except for the number of singlet fields.
The relevant superpotential is given by (\ref{W})-(\ref{Wc}).
The essential difference is that
the VEV of the diagonal component of $A$
points to the direction $5V'-9V+24Y$.

Because the contribution to mass of $\Phi$ comes only from 
$W_{\bar\Phi'}$,
we concentrate on this interaction.
The component fields which have the same $U(1)$ charges as ${\bf 1}_\Phi$ are
${\bar L}_{\bf 10}$, $L_{\bf 16}$, and $E_{\bf 16}$, and they are massless
by the sliding singlet mechanism, though the component fields
$L_{\bf 16}$ and $E_{\bf 16}$ are absorbed by the Higgs mechanism. 
The component field which has the same $U(1)$ charge as ${\bf 16}_\Phi$
is $L_{\bf 10}$, so we have to examine whether the sliding singlet mechanism
acts for ${\bf 16}_\Phi$ in this model. 
In the $F$-flatness condition 
\begin{equation}
F_{{\bf\overline{16}}_{\Phi'}}=(1+A+Z_i+A^2+AZ_i+Z_i^2){\bf16}_\Phi
   +(1+A+Z_i){\bf16}_A{\bf1}_\Phi=0, 
\end{equation}
the coefficient of the first term does not vanish because the component
${\bf16}_\Phi$ has the different charge from ${\bf 1}_\Phi$. 
Therefore, the above $F$-flatness condition means just that 
the sum of the two terms must vanish, namely, the sliding singlet mechanism
does not act for the component ${\bf16}_\Phi$. 
The remaining task is to check that there are no additional massless 
modes, which can be done straightforwardly, and we skip it here. 
In this model, the massless Higgs $\bar L$ comes from ${\bf10}_\Phi$ and 
the main mode of massless Higgs $L$ comes from a primed field.

Unfortunately we have not found a realistic quark and lepton sector 
compatible with this model, in contrast to the model in \S\ref{e6dw}. 
This is because, in the context of anomalous $U(1)_A$, primed fields,
which have positive anomalous $U(1)_A$ charges,
tend to have suppressed coupling constant, and therefore the bottom 
and $\tau$ Yukawa couplings become too small. 
In order to avoid this, the massless doublet Higgs should belong to 
unprimed fields. 
However, it is difficult to construct such a model without extra 
massless modes, in this breaking pattern. 
One of the reason is that the sliding singlet mechanism acts for the 
$L$ mode in 
${\bf 16}_\Phi$ which is absorbed by the Higgs mechanism. 
To avoid this situation, we can change the direction of $U(1)$ so that
the sliding singlet mechanism acts to $L$ and $\bar L$ modes in 
${\bf 10}_\Phi$, but this vacuum is nothing but that of the DW type.
It is another possibility that the massless Higgs $L$ comes from 
${\bf \overline{16}}_{\bar C}$. To realize this situation, we can apply the
sliding singlet to the $\bar C$ field. However, in that situation, there 
appear several undesired massless modes, which spoils the success of the
gauge coupling unification.

\subsection{Model IV: $SO(10)$\sub{F}$\times U(1)$}

For this breaking pattern, even if the sliding singlet mechanism 
acts well, the DT splitting can not be realized, 
although the doublet Higgs indeed become massless.
This is because the triplet Higgs in {\bf16}$\in${\bf27} 
also become massless.
In terms of $SO(10)$\sub{F}$\times U(1)$, it is possible 
to give a large mass to the triplet Higgs while the doublet remains 
massless, through the missing partner mechanism similar as 
$SU(5)$\sub{F}$\times U(1)$\cite{flippedsu5} case.
However, it is difficult to embed this 
$SO(10)$\sub{F}$\times U(1)$ model into a \E6 model. 
This topic is discussed in detail in Ref.\cite{flippedso10}.

\section{Summary and Discussion}

In this paper, we extracted the essence of the sliding singlet
mechanism in which SUSY breaking effect does not spoil the doublet-triplet
splitting.
And we generalized the sliding singlet mechanism.
The essential point in this mechanism
is that the sliding singlet mechanism makes the doublet components massless 
which have the same $U(1)$ charges as the SM singlet components which have
non-vanishing VEVs.%
\footnote{Conversely, components with different charges from the SM singlet 
component become massive, that has been used to make pseudo NG modes massive
in the literature\cite{BarrRaby}.
}
By choosing the $U(1)$ which is determined by the VEV
of the adjoint Higgs field, we can build various GUTs in which the generalized
sliding singlet mechanism acts. In this paper, we examined various $E_6$ GUTs
by using anomalous $U(1)_A$ gauge symmetry, by which we can 
construct GUT models with generic interactions (including even higher
dimensional interactions). Since in $E_6$ group there are three $U(1)$ which
commute with the SM gauge group, we can select two of three doublets 
in the fundamental representation ${\bf 27}$ to become massless
by the sliding singlet mechanism, though one of the three is enough to realize
the DT splitting. 
Among the various GUTs we examined in this paper, it is the most promising
model in which the $L$ and $\bar L$ components in ${\bf 10}$ of $SO(10)$ in 
${\bf 27}$ of $E_6$ become massless by the sliding singlet mechanism. 
This vacuum is nothing but the Dimopoulos-Wilczek type vacuum.
To make the $\bar L$ component in ${\bf 27}$ massless by the sliding singlet
mechanism is important to realize large top Yukawa coupling. For this purpose,
it is enough that $\bar L$ in ${\bf 27}$ becomes massless by the sliding 
singlet mechanism, namely, the charge of ${\bar L}$ is taken to be the same
as that of the SM singlet in ${\bf 27}$. The concrete condition 
is that $(a,b,c)=(\frac{1}{4},-\frac{3}{4}+\frac{3}{2}c,c)$ in the notation
in Table \ref{qe6}.
Because $L$ component in ${\bf 16}$ is absorbed by the Higgs mechanism, 
it is better that the other $L$ component in ${\bf 10}$ becomes massless 
independently by some mechanism. 
Of course, to realize DT splitting, it is enough that $\bar L$ component 
Higgs is guaranteed to be massless, because there must be its massless 
partner $L$.
However, in that case, the main component of the partner $L$ tends 
to come from positively anomalous $U(1)_A$ charged field (primed field), 
because positively
charged fields have smaller masses than negatively charged fields. 
And positively charged Higgs $L$ leads to small down-type quark Yukawa 
couplings and therefore too small $\tan \beta$. 

The generalized sliding singlet mechanism has opened the new possibility
to build various GUT models in which DT splitting is realized. The DW 
type vacuum is the most promising vacuum in $E_6$ GUTs even in the sense of
the sliding singlet mechanism, though the other possibilities may become also
interesting. We hope that such an observation gives us a key leading 
to the real GUT which describes our world.

\section*{Acknowledgement}
N.M. is supported in part by Grants-in-Aid for Scientific 
Research from the Ministry of Education, Culture, Sports, Science 
and Technology of Japan.

\end{document}